\documentclass[aps,prb,amsmath,twocolumn,amssymb,floatfixng,showpacs,
superscriptaddress,footinbib]{revtex4-1}
\pdfoutput=1
\usepackage[dvips]{graphics}
\usepackage{bm}
\usepackage{float}
\usepackage{epsfig}
\usepackage{enumerate}
\usepackage{amsmath}
\usepackage{color}
\usepackage{graphicx}
\usepackage[colorlinks=true,linktoc=page,linkcolor=magenta,citecolor=blue]{hyperref}

\newcommand{\ie}{{\it i.e.,}}

\newcommand\bea{\begin{eqnarray}}
\newcommand\eea{\end{eqnarray}}
\newcommand\beq{\begin{equation}}  
\newcommand\eeq{\end{equation}}

\begin{document} 
\title{Jackiw-Rebbi zero modes in non-uniform topological insulator nanowire} 

\author{Sayan Jana}
\email{sayan@iopb.res.in}
\affiliation{Institute of Physics, Sachivalaya Marg, Bhubaneswar-751005, India}
\affiliation{Homi Bhabha National Institute, Training School Complex, Anushakti Nagar, Mumbai 400085, India}
\author{Arijit Saha}
\email{arijit@iopb.res.in}
\affiliation{Institute of Physics, Sachivalaya Marg, Bhubaneswar-751005, India}
\affiliation{Homi Bhabha National Institute, Training School Complex, Anushakti Nagar, Mumbai 400085, India}
\author{Sourin Das}
\email{sdas.du@gmail.com}
\affiliation{Department of Physical Sciences, IISER Kolkata, Mohanpur, West Bengal 741246, India}

\begin{abstract}
We theoretically investigate the emergence of Jackiw-Rebbi zero modes and their conductance signature in non-uniform topological insulator nano-wires. We modelled the non-uniform nano-wires as junction between two cylindrical nano-wires with different radius. In the limit of wire length being much larger than its radius, the surface state of the nanowire splits into one dimensional Dirac modes propagating along the axis of the cylinder owing to radial confinement. The sign of the mass gap in each of these Dirac mode is decided by angular momentum quantum number corresponding to the rotational motion of the electron about the axis of the cylindrical. Application of an external magnetic flux through the cylindrical nanowires enables us to tune the mass gap from positive to negative value across the junction.  
Due to this flux tunable band inversion, controlled by the external magnetic filed, Jackiw-Rebbi zero modes can be made to appear or disappear at the junction. We compute differential conductance of our topological insulator nanowire junction and show that a quantized conductance peak appears at zero-energy (zero-bias) in the presence of the Jackiw-Rebbi mode. 
\end{abstract}

\maketitle
{\it{Introduction:-}}
In recent years, the search for exotic phases like fractional charge excitations in modern condensed matter systems, has drawn a large attraction owing to their rich fundamental physics. Also such excitations are believed to be a possible candidate for performing topological quantum computation~\cite{nayak2008non}. One particular example of exotic excitation with fraction charge is the fractional fermions (FFs) or Jackiw-Rebbi modes (JRMs)~\cite{jackiw1976solitons} where it was shown that one-dimensional (1D) Dirac field coupled to a static background soliton field gives rise to fractional charge excitations~\cite{RajaramanBell1982, Rajaraman2001} localized about the mass domain wall. In the seminal work by Su, Schrieffer and Heeger (SSH)~\cite{su1980solitonPRL, su1980soliton}, they predicted the existence of similar JRMs and their topological nature and applied it to describe the physics of polyacetylene. In recent times, several theoretical proposals have been made to realize JRMs in various condensed matter systems. In particular, they can arise in finite-length nanowires (NWs), quantum dots and rings in the presence of a charge-density wave gap induced by a periodic modulation of the chemical potential~\cite{gangadharaiah2012localized, PhysRevB.94.075416}, 1D zigzag fermion chain~\cite{PhysRevLett.107.166804} and in quantum spin Hall state~\cite{qi2008fractional, PhysRevB.94.241406}. Moreover, very recently, it was shown that simultaneous presence of Rashba spin-orbit interaction (SOI) along with uniform and spatially periodic magnetic field in 1D NWs can support gapped phases hosting zero modes of Jackiw-Rebbi type~\cite{klinovaja2012transition, rainis2014transport, JelenaLoss} obeying non-abelian statistics~\cite{PhysRevLett.110.126402}. Theoretical proposals for transport signature of JRMs have been put forwarded in Refs.~[\onlinecite{rainis2014transport}, \onlinecite{saha2014quantum}]. However, till date, no direct experimental evidence of JRMs in any of these above recently discussed systems 
has been reported so far. 

The recent discovery of a new class of materials called topological insulators (TIs)~\cite{kane2005quantum, bernevig2006quantum, moore2010birth, hasan2010colloquium, qi2011topological, konig2007quantum, culcer2012transport, xia2009observation} has accelerated the search for new exotic zero energy modes, in particular Majorana fermions 
(MFs)~\cite{fu2008superconducting, alicea2012new}. These modes are topologically protected by the inherent symmetry of the system and therefore robust against any local perturbation. 
Very recently, the emergence of MFs has been theoretically predicted in three dimensional topological insulator (3D TI) nanowire (NW) in presence of proximity induced 
superconductivity~\cite{cook2011majorana, roni2014PRL, GYHuangPRB2017, de2018conditions}. 3D TI NW with perfectly insulating bulk can be modelled as an ideal hollow metallic cylinder with 
a diameter large enough such that it is amenable to thread a large magnetic flux through its core ~\cite{ran2008spin, rosenberg2010wormhole, de2018conditions, cook2011majorana, roni2014PRL, 
PhysRevLett.108.106403, MESAROS2013977}. Few transport properties like Josephson current~\cite{RoniIlanNJP2014}, magneto conductance~\cite{moors2018magnetotransport}, 
thermoelectric currect~\cite{thermoelectricTINW2017} etc. have been investigated in the context of 3D TI NW. In recent times, quasi-ballistic 3D TI NWs have been experimentally realized 
in order to investigate Aharonov-Bohm oscillations~\cite{cho2015aharonov}, Josephson current~\cite{APL2018} etc. via them. Nevertheless, prediction of zero energy JRMs and their 
transport signal, in the context of 3D TI NW has not been put forwarded so far to the best of our knowledge. 

\begin{figure}[!thpb]
\centering
\includegraphics[width=1. \linewidth]{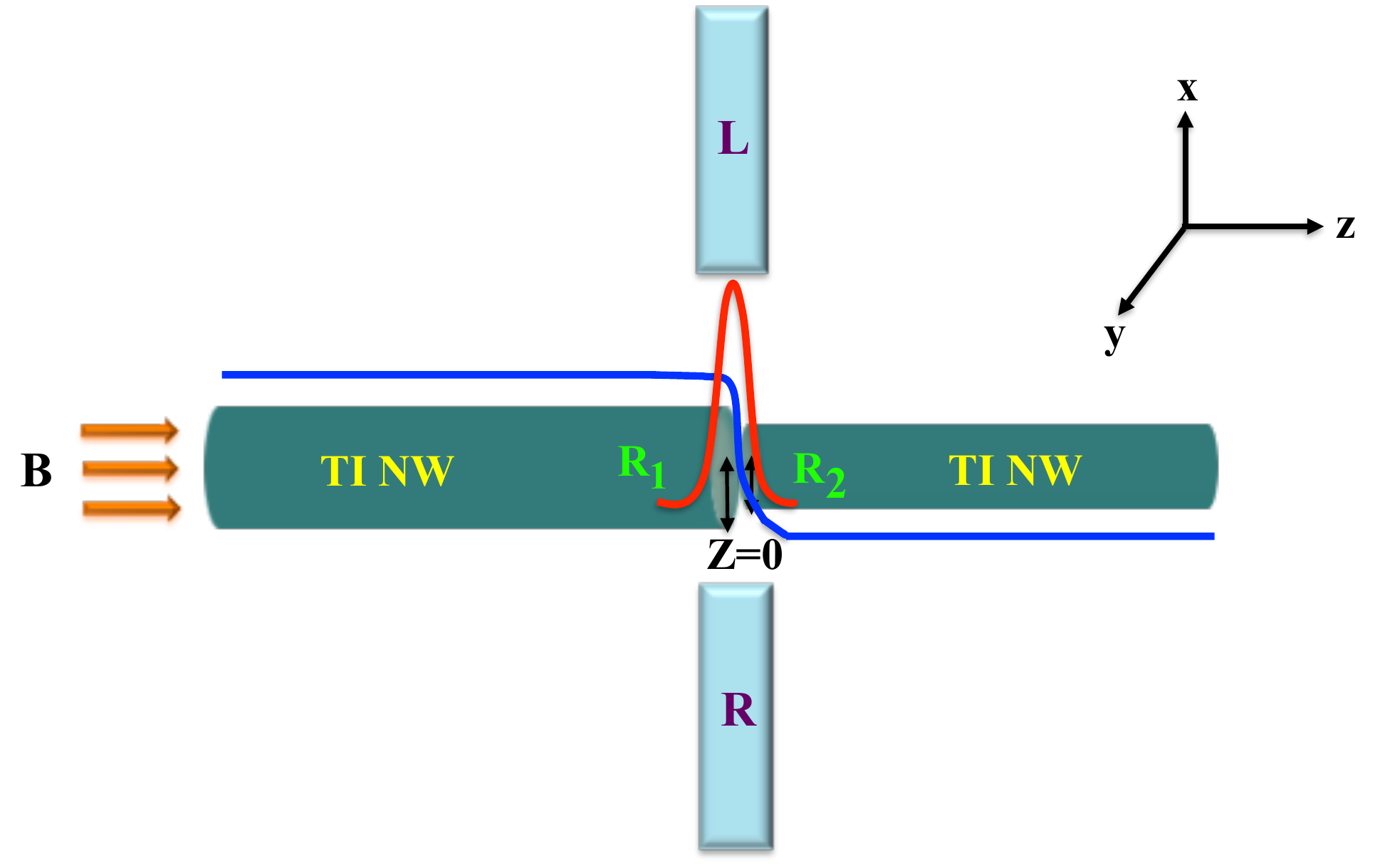}
\caption{(Color online) Schematic of our set-up in which two cylindrical shaped 3D TI NWs with different radii ($R_{1}$ and $R_{2}$) form a junction. The axis of both the NWs are chosen along 
the $z$ direction. An external magnetic field $B$ has been incorporated along the axis of both the NWs. The sign change of mass gap and the appearence of zero energy Jackiw-Rebbi mode 
at the junction ($z=0$), are denoted by blue (grey) and red (grey) solid lines respectively. Two leads are placed on the two sides of the junction to attain dominant contribution appearing from the 
Jackiw-Rebbi zero mode such that the differential conductance reaches a maximum value of $e^2/h$ (see text for discussion).}
\label{fig1}
\end{figure}

In this article, we consider a junction of two 3D TI NWs with different radii and show that one can realize zero energy Jackiw-Rebbi mode at the junction of two NWs. Within our geometry, in the limit of wire length being much larger then its radius, the surface state of the nanowire splits into one dimensional Dirac modes owing to radial confinement. The sign of the mass gap in each of these Dirac modes is then tuned by application of an external magnetic flux through the cylindrical. This enables a possibility of flux tunable mass gap inversion across a junction of two TI NWs with different radius leading to emergence of Jackiw-Rebbi zero modes at the junction. We study the differential conductance of our TI NWs junction by employing the scattering matrix approach ($\rm Weidenm\ddot{u}ller$ formulae)~\cite{fisher1981relation,iida1990statistical,haim2015signatures} and show that a zero-bias peak (ZBP) appears in the conductance spectrum. Such ZBP can be a possible transport signal of the 
JRMs present in our set-up. 

The outline of this article is organized as follows. First, we describe our model and energy spectrum of a 3D TI NW. Then, we discuss our results for appearence of JRMs due to the mass
inversion accross the junction of two cylindrical shaped TI NWs (both non-twisted and twisted case) and their corresponding differential conductance signature. Finally, we summarize and conclude.

{\it{Model Hamiltonian and band spectrum:-}}
We briefly review the surface state Hamiltonian of a 3D TI~\cite{cook2011majorana, rosenberg2010wormhole} which can be written as,
\begin{eqnarray}
&&H_{TI}=\frac{v}{2}[\hslash\nabla.\bar{n}+\hat{n}.(\vec{P}\times\vec{\sigma})+\hat{n}.(\vec{\sigma}\times\vec{P})]\ .
\label{equ1}
\end{eqnarray}
where $v$ is the velocity of the surface electrons. $\vec{P}=-i\hslash\nabla$ is the quantum mechanical momentum operator for electrons, $\vec{•\sigma}=(\sigma_{x},\sigma_{y},\sigma_{z})$ 
is the Pauli spin matrices and $\hat{n}$ denotes the unit vector normal to the cylindrical surface. Considering the cylindrical axis along $\hat{z}$ direction and $\hat{n}=\hat{r}$, one can obtain 
the TI NW (schematically shown in Fig.~\ref{fig1}) Hamiltonian in matrix form as~\cite{de2018conditions},
\begin{eqnarray}
&&H_{TI}=\Big[-iv\hslash\Big(\frac{\sigma_{z}\partial_{\theta}}{R}-\sigma_{y}\partial_{z}\Big)-\mu\Big]\ ,
\label{eq2}
\end{eqnarray}
Here, we have used the two degrees of freedom for the cylindrical surface, the coordinate along the axis of the cylinder given by $z$ and the radial degree of  freedom about the $z$-axis given by $\theta$.
The energy eigen values corresponding to this Hamiltonian are given by 
\begin{eqnarray}
&& E_{l\pm}(k)=\pm v\hslash\sqrt{\frac{l^2}{R^2}+k^2}-\mu\ ,
\label{eq4}
\end{eqnarray}
where $R$ is the fixed radius of the cylindrical TI NW and $\mu$ is the chemical potential.  From Eq.(\ref{eq4}) and Eq.(\ref{eq2}), one can conclude that, this energy spectrum is a spectrum of a 1D Dirac fermion propagating along the $z$ direction with a mass gap given by $l/R$ where $l$ is the angular momentum quantum number. 

\begin{figure}[!thpb]
\centering
\includegraphics[width=0.75 \linewidth]{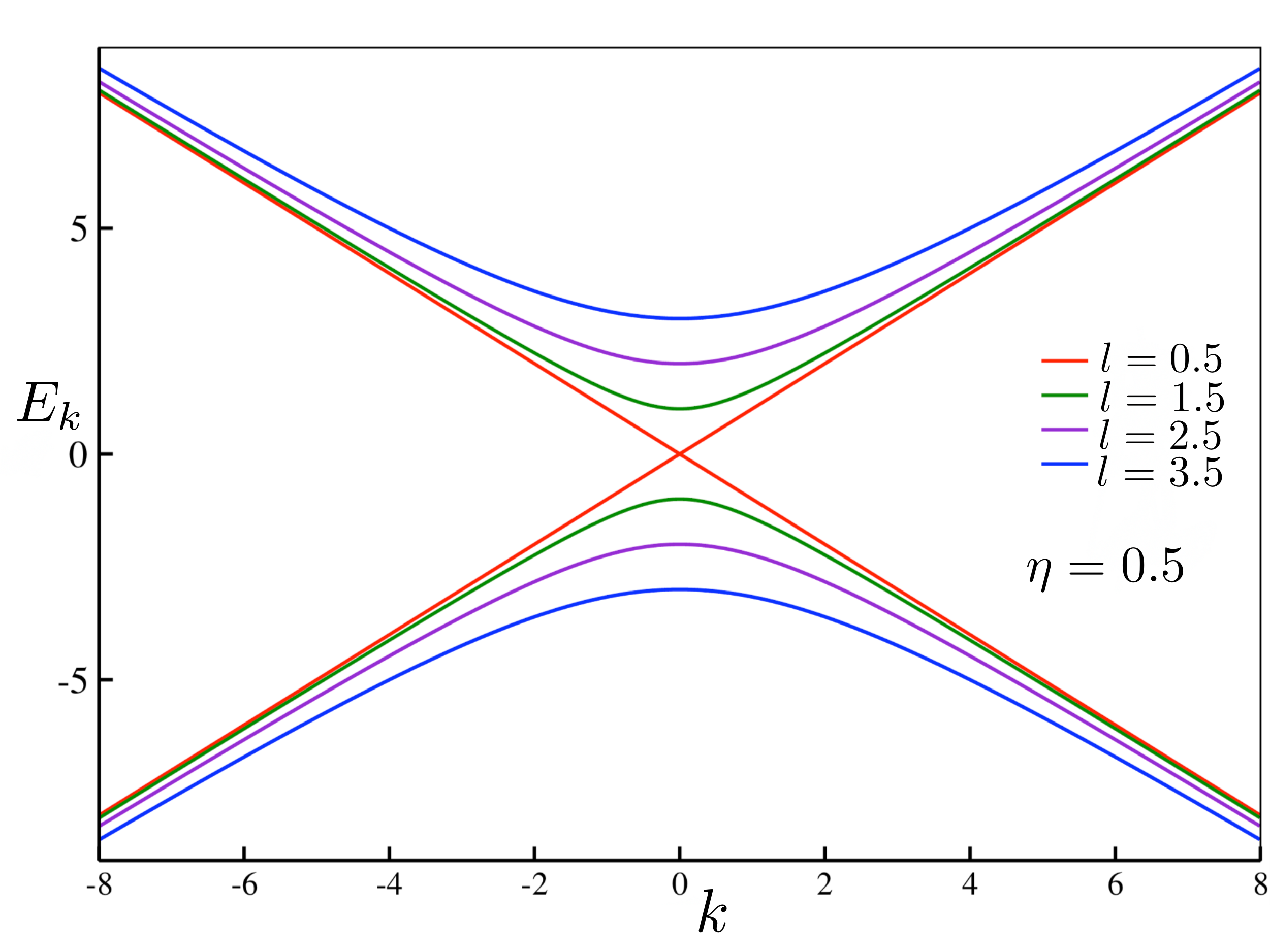}
\caption{(Color online) Energy spectrum of a 3D TI NW is depicted for different angular momentum modes $l$.}
\label{fig2}
\end{figure}

We employ an anti-periodic boundary condition of the wave function of the TI NW in the angular coordinate $\theta$. Hence the angular momentum quantum numbers are given by half integer \ie~$l=\pm\frac{1}{2}, \pm \frac{3}{2}, \pm\frac{5}{2}, \pm\frac{7}{2}$ etc. $k$ is the linear momentum along the cylinder axis 
($z$ axis in our analysis). Now, by applying an external magnetic field $B$ through the cylinder axis leads to a magnetic flux $\Phi$ piercing the core of the cylinder. This can be included in the Hamiltonian by replacing the momentum operator with $\vec{\Pi}$=($\vec{P}$ - $\frac{e{\vec{A}}}{c}$)  where ${\vec{A}}= \frac{\eta\phi_{0}\hat{z}\times\vec{r}}{2\pi r^2}$ is the vector potential. 
Therefore, $\Phi=\eta\Phi_{0}$ represents the total magnetic flux piercing the cylinder.

Upon incorporating the magnetic field, the angular momentum operator translates by $-i\partial_{\theta}\rightarrow $($-i\partial_{\theta}$ - $\frac{\Phi}{\Phi_{0}}$), where $\Phi_{0}=hc/e $ is the flux quanta.
Hence, the energy spectrum becomes~\cite{cook2011majorana}
\begin{eqnarray}
&& E_{l\pm}(k)=\pm v\hslash\sqrt{\frac{(l-\eta)^2}{R^2}+k^2}-\mu\ .
\end{eqnarray}
The corresponding band spectrum is demonstrated in Fig.~\ref{fig2}. Note that, at $k=0$ this spectrum is associated with a band gap which depends on $(l - \eta)$ for fixed radius $R$ of the cylinder. Also, the gap in the surface state spectrum vanishes if one considers the limit $R \rightarrow \infty$ as expected. It is also evident that one can control the closing and opening of this gap in the spectra by modulating 
the flux through the cylinder. 

For example, for the case of ($\eta=\frac{1}{2},l=\frac{1}{2}$), the gap at $k=0$ vanishes implying a massless Dirac spectrum. This essentially implies that the Dirac fermions corresponding to the $l=\frac{1}{2}$ undergoes a mass inversion (band inversion) as $\eta$ varies form $\eta=\frac{1}{2}-{\tilde{\delta}}$ to  $\eta=\frac{1}{2}+{\tilde{\delta}}$ where ${\tilde{\delta}}$ is a small positive number. One  can physically interpret the gap closing in terms of Berry phase. When $\eta=\frac{1}{2}$, the Berry phase accumulated due to the rotations of the spins constrained to move on the surface of the cylinder, is nullified by the Aharonov-Bohm phase due to the flux. This preserves the time reversal symmetry. 

{\it{Results:-}}
In the seminal work by Jackiw and Rebbi, they showed  that Dirac field coupled to background soliton scaler field gives rise to exotic zero modes ~\cite{jackiw1976solitons} associated with charge fractionalization~\cite{RajaramanBell1982, Rajaraman2001}. To realize such zero energy modes in TI NWs, we consider a junction of two cylindrical 3D TI NWs with radius $R_{1}$ and $R_{2}$. Our set-up is schematically depicted in Fig.~\ref{fig1}. Both the NWs lie along the $z$ direction \ie~we consider $z$ axis as cylinder axis. A magnetic field $B$ is applied along the axis of the NWs to induce magnetic flux through the core of them. 

Now for the regime $z<0$, we consider the radius of the NW to be $R=R_{1}$ and chemical potential $\mu$=$\mu_{1}$. Hence, our Hamiltonian for $z<0$ becomes
 \begin{eqnarray}
 &&H_{TI}(z<0)=\Bigg[-v\Bigg(\frac{\sigma_{z}(i\partial_{\theta}+\eta_{1})}{R_{1}}-i\sigma_{y}\partial_{z}\Bigg)-\mu_{1} \Bigg]\ ,
  \label{eq11}
 \end{eqnarray}
Similarly for the regime $z>0$, considering a NW with radius $R=R_{2}$ and chemical potential $\mu$=$\mu_{2}$, our Hamiltonian can be written as
\begin{eqnarray}
 && H_{TI}(z>0)=\Bigg[-v\Bigg(\frac{\sigma_{z}(i\partial_{\theta}+\eta_{2})}{R_{2}}-i\sigma_{y}\partial_{z}\Bigg)-\mu_{2}\Bigg]\ ,
 \label{eq12}
 \end{eqnarray}
Here, $\eta_{1}=\pi B R_{1}^{2}/\Phi_{0}$ and $\eta_{2}=\pi B R_{2}^{2}/\Phi_{0}$ are the fluxes through the two cylindrical TI NWs respectively.

In order to have a normalizable bound state solution near the junction ($z=0$) the wave function must vanish at $z=\pm\infty$. 
Also the wave function must be continuous across the boundary. The choice of trial wave function that vanishes at $z=-\infty$ reads as
\begin{eqnarray}
&&\Psi_{z<0}=
   \begin{bmatrix} 
   \phi_{1-}   \\
   \phi_{2-} \\
 \end{bmatrix} 
 \exp({\lambda_{-} z+i l_{1}\theta})\ ,
\end{eqnarray}
where, from Eq.(\ref{eq11}) we have $\lambda_{-}=\sqrt{\frac{(l_{1}-\eta_{1})^2}{R_{1}^2}-\frac{(\mu_{1}+E)^2}{v^2}}$.
Similarly the choice of trial wave function that vanishes at $z=+\infty$ can be written as
\begin{eqnarray}
&&\Psi_{z>0}=
   \begin{bmatrix} 
   \phi_{1+}   \\
   \phi_{2+} \\
\end{bmatrix} 
  \exp({-\lambda_{+} z+i l_{2}\theta})\ ,
\end{eqnarray}
where $\lambda_{+}=\sqrt{\frac{(l_{2}-\eta_{2})^2}{R_{2}^2}-\frac{(\mu_{2}+E)^2}{v^2}}$ from Eq.(\ref{eq12}).

Because of the continuity condition at the boundary $z=0$, the wave function requires $\Psi_{z<0}(z, \theta)=\Psi_{z>0}(z, \theta)$. 
Additionally we consider a junction which respects rotational symmetry so that we have  $l_{1}=l_{2}=l$. 
We obtain an expression for the energy of the bound state as
\begin{eqnarray}
&&E=\Bigg[\frac{-\mu_{1}[(l_{2}-\eta_{2})R_{1}] + \mu_{2}[(l_{1}-\eta_{1})R_{2}]}{(l_{2}-\eta_{2})R_{1}-(l_{1}-\eta_{1})R_{2}}\Bigg]\ ,
\label{eq14}
\end{eqnarray}
 In Eq.(\ref{eq14}), there exists a solution for $E=0$ with the condition
$(l_{1}-\eta_{1})=-(l_{2}-\eta_{2})=\delta$ and $\mu_{1}R_{1}=-\mu_{2}R_{2}=\gamma$. From this condition $(l_{1}-\eta_{1})=-(l_{2}-\eta_{2})$, it is evident that the presence of the zero energy solution is directly related to the inversion of mass gaps across the junction where the band gap for the respective NWs are proportional to $(l_{1}-\eta_{1})$ and $ (l_{2}-\eta_{2})$ respectively. One can continuously change the mass gaps and induce  mass inversion by controlling the magnetic field as leading to JRMs for $l_1=l_2=1/2, 3/2$ etc. 

\begin{figure}[!thpb]
\vskip +0.4cm
\centering
\includegraphics[width=1.05 \linewidth]{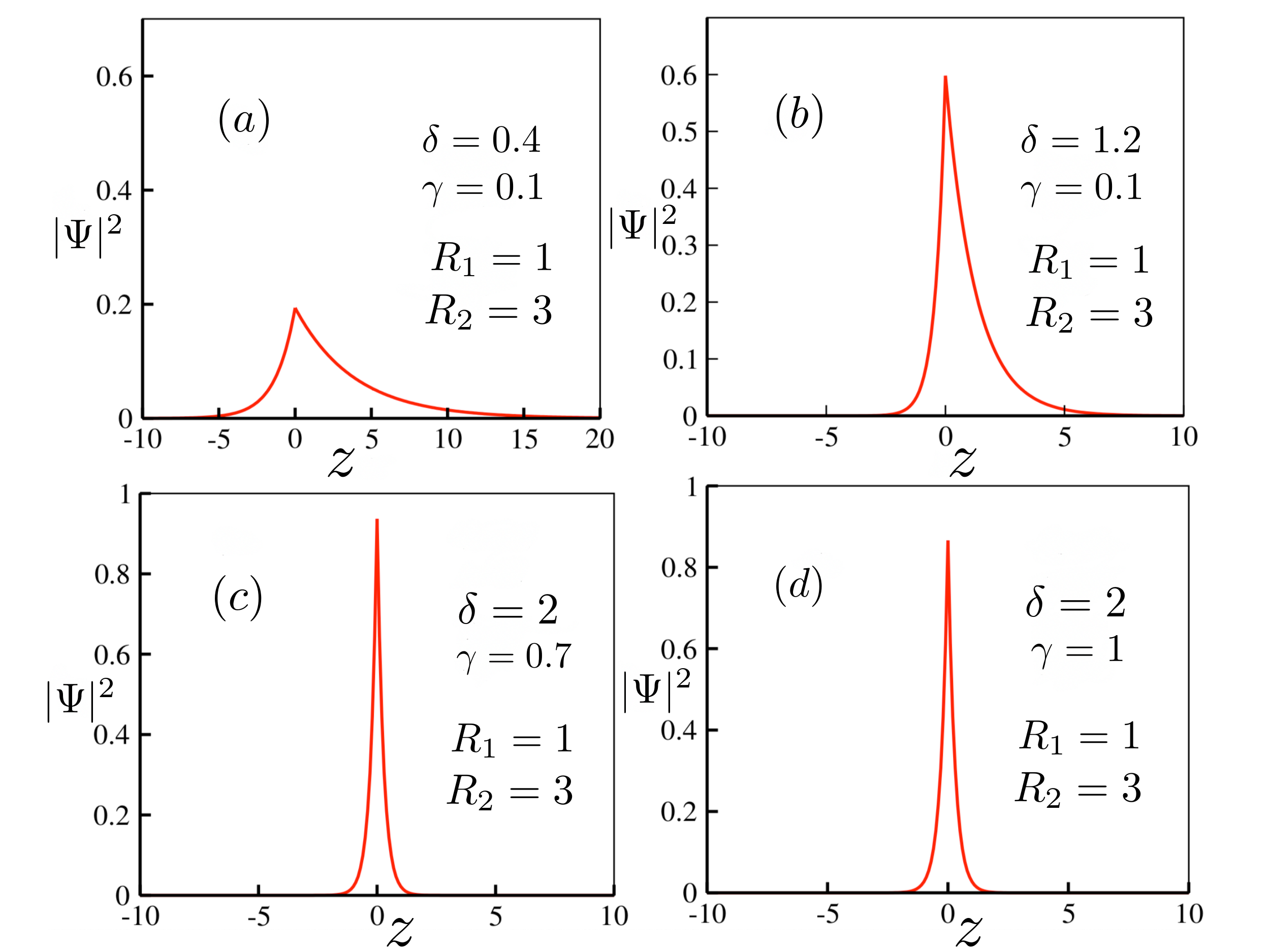}
\caption{(Color online) Asymmetric probability density $\lvert \Psi \rvert ^{2}$ of the Jackiw-Rebbi zero energy modes are illustrated as a function of distance $z$ for different values of $l$.}
\label{fig5}
\end{figure}

In Fig.~\ref{fig5}, we show the behavior of probability density $\lvert \Psi \rvert ^{2}$ corresponding to the zero mode ($E=0$) wave function as a function of distance $z$ from the junction 
for various parameter values. Note that, for smaller values of $l$, the zero-modes are asymmetrically spread around the junction. With the enhancement of the $l$ values, the zero mode gradually becomes more and more localized at the junction owing to the increased mass gap in the spectrum.  

Furthermore, we consider two cylindrical TI NWs junction as previously shown in Fig.~\ref{fig1} which are rotated with respect to each other about the $z$-axis. To be specific, we rotate the $y$ component 
of the spin ($\sigma_{y}$) to $x$ component ($\sigma_{x}$) 
in the right TI NW ($z>0$). Hence, $\sigma_{x}$ is now coupled to the linear momentum $k$ along the positive $z$ direction. This mimics a situation in which the spin of the electrons is rotated while passing through the junction due to the presence of non-uniformity of spin-orbit coupling across the junction. 

Therefore, one can write down the Hamiltonian of the two TI NWs with two different spin components in a compact form as~\cite{ilan2012nonequilibrium},
\begin{eqnarray}
&&H_{TI}(z)=-iv\hslash\frac{\sigma_{z}\partial_{\theta}}{R(z)}+\frac{iv\hslash}{2} \sigma_{x}\big \{ \cos\theta(z),\partial_{z} \big \}\nonumber \\
&&+\frac{iv\hslash}{2} \sigma_{y}\big \{ \sin\theta(z),\partial_{z} \big \}-\mu(z)\ ,
\label{combH}
\end{eqnarray}
where, for $z<0$, $\theta(z)=0$, $R(z)=R_{1}$ and $z>0$, $\theta(z)=\pi/2$, $R(z)=R_{2}$.

One has to solve the $\rm Schr\ddot{o}dinger$ equation $ H\Psi=E\Psi$ to obtain the correct matching condition at the boundary ($z=0$) of the two cylindrical TI NWs. This can be
done by finding a proper transfer matrix which correctly matches the wave-functions for $z<0$ and $z>0$ at the boundary as
\begin{eqnarray}
&&\Psi_{z>0}=T_{(z>0,z<0)}\Psi_{z<0}\ ,
\end{eqnarray}
Following Ref.~[\onlinecite{ilan2012nonequilibrium}], we derive the transfer matrix for our case as,
\begin{widetext}
\vskip -0.5cm
\begin{align}
&&T_{(z,-z)}=P_{x}\Bigg[\exp{\int_{-z}^{z} dz \bigg \{\frac{-v\hslash\sigma_{z}l/R(z)+E+\mu(z)+\frac{iv\hslash}{2}(\sigma_{x}\sin\theta(z)-\sigma_{y}\cos\theta(z))\partial_{z}\theta(z)}{M} \bigg \}}\Bigg]\ ,
\label{eq22}
\end{align}
\end{widetext}
where the matrix $M$ in Eq.(\ref{eq22}) can be written as
\begin{eqnarray}
&&M=iv\hslash(\sigma_{y}\sin\theta(z)+\sigma_{x}\cos\theta(z))\ .
\end{eqnarray}

Therefore, employing the appropriate step function in $\theta(z)$ and considering $z\rightarrow{\tilde\alpha}$ (${\tilde\alpha}$ is a small positive number) and ${\tilde\alpha}\rightarrow 0$ at the end, 
we solve Eq.(\ref{eq22}) and obtain the wave-function matching condition at the TI NWs bounday as
\begin{equation}
\Psi(0^{+})=e^{-(i\sigma_{z}\pi)/4}\Psi(0^{-})\ ,
\end{equation}
Then computing the wave functions by solving the Hamiltonian in Eq.(\ref{combH}) and following the same procedure as described before, we obtain the same condition 
for the zero mode solution. This manifests the robustness of the Jackiw-Rebbi zero modes with the rotation of the electron spin accross the boundary.

 \begin{figure*}[!bthp]
\hspace*{\fill}%
\includegraphics[width=0.5 \linewidth]{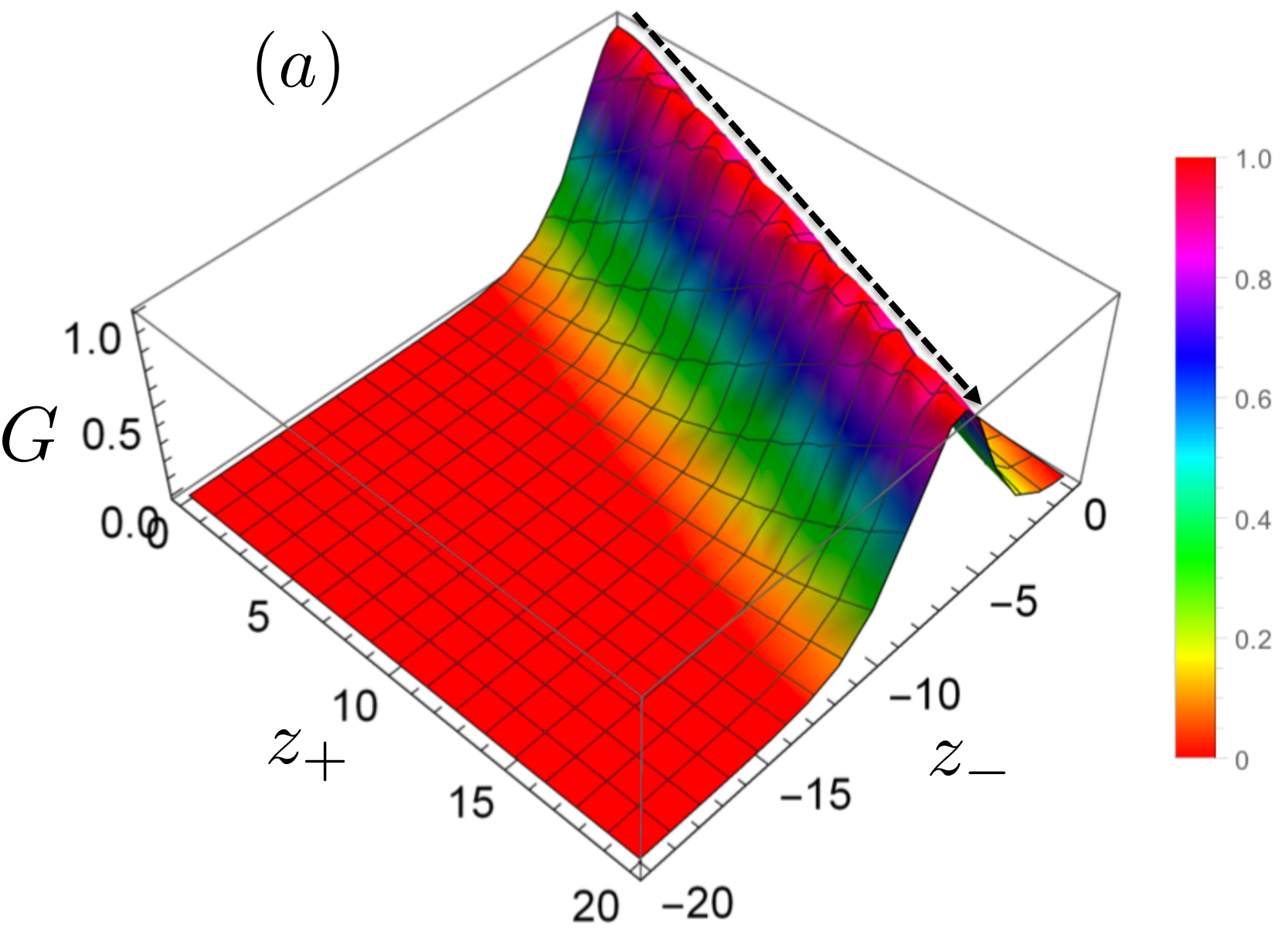}\hfill%
\includegraphics[width=0.5 \linewidth]{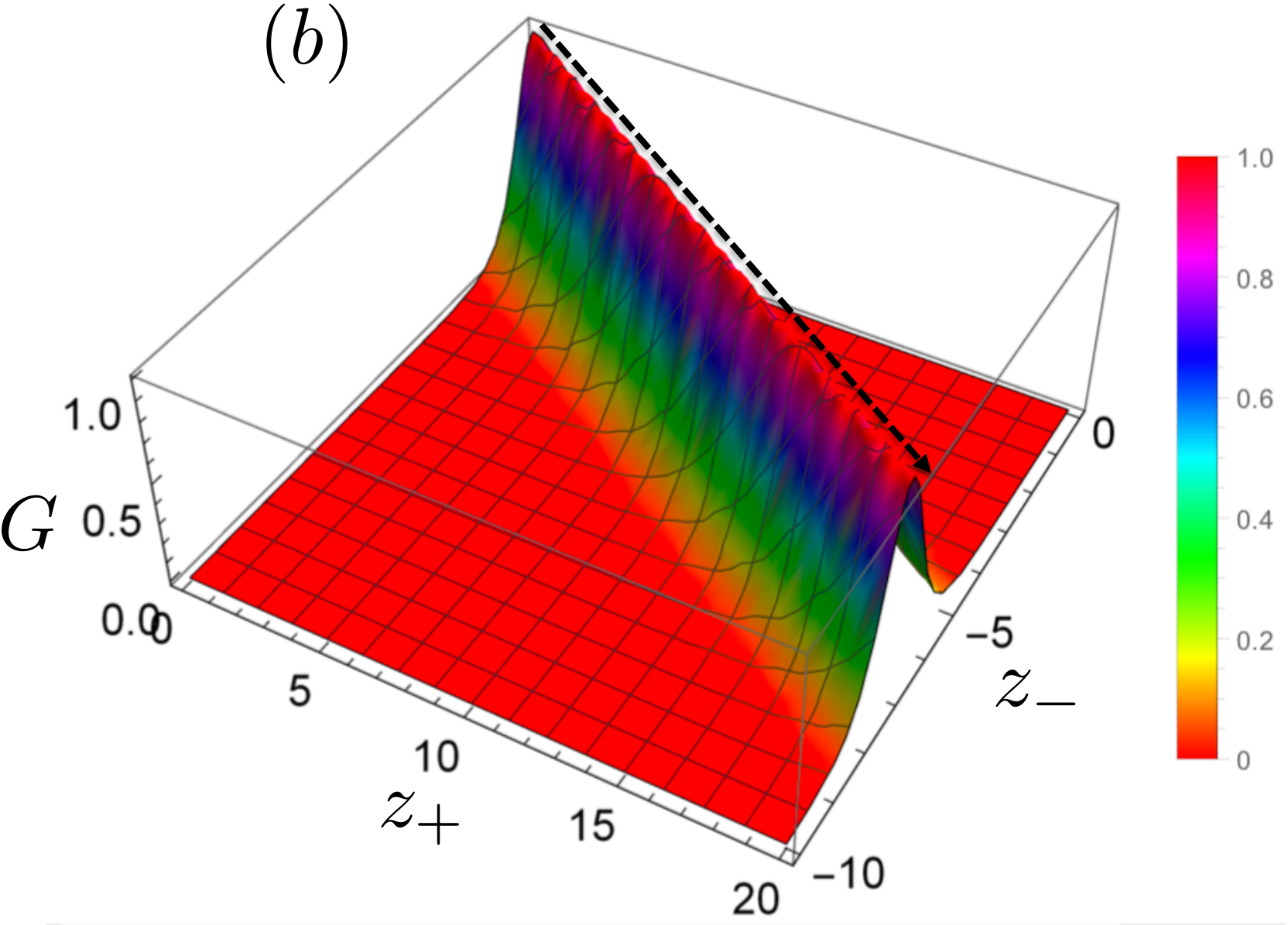}
\hspace*{\fill}%
\caption{(Color online) Zero energy ($E=0$) differential conductance spectra, via the Jackiw-Rebbi zero modes, is demonstrated (in units of $e^{2}/h$) as a function of distance from the junction. 
The value of the other parameters are chosen for panel (a) $\delta=0.4$, $\gamma=0.1$, $R_{1}=1$, $R_{2}=3$ and panel (b) $\delta=1.2$, $\gamma=0.1$, $R_{1}=1$, $R_{2}=3$.}
\label{fig6}
\end{figure*}
 
Finally, we discuss the differential conductance signal of the Jackiw-Rebbi zero modes present in our TI NWs set-up. We consider a situation where a tunnel probe (leads depicted in Fig.~~\ref{fig1}) 
is in connection to each side of the junction such that the conductance between the probes connected to the left and right side of the junction is dominated by the contribution arising from the zero modes. 
The local tunneling Hamiltonian describing this situation can be  written as

\begin{eqnarray}
H_{T}=\tilde{\gamma}^{\dagger}\sum_{\alpha,k}^{} u_{\alpha}c_{\alpha k}+h.c\ ,
\label{eq26}
\end{eqnarray}

Here, $\alpha=L, R$ denotes the left and the right leads, $\tilde{\gamma}^{\dagger}$ and $c_{\alpha k}$ correspond to the creation operator for the zero mode and annihilation operator of the electrons  in the leads respectively. We compute the scattering amplitudes employing $\rm Weidenm\ddot{u}ller$ formula~\cite{fisher1981relation,iida1990statistical,haim2015signatures} and they read as

\begin{eqnarray}
&&r_{\alpha}=1-\frac{2 i u_{\alpha} \pi u_{\alpha}^{\star}}{E+i \pi (\sum_{\alpha}u_{\alpha}^{2})}\ ,
\label{eqr}
\end{eqnarray}
\begin{eqnarray}
&&t_{\alpha}=-\frac{2 i u_{\beta} \pi u_{\alpha}^{\star}}{E+i \pi (\sum_{\alpha}u_{\alpha}^{2})}\ ,
\label{eqt}
\end{eqnarray}

Therefore, considering the lead wave functions as plane wave states ($\Phi_{\alpha}$), the tunneling amplitudes $ u_{\alpha} $ in Eqs.(\ref{eqr}-\ref{eqt})  represent the  overlap of the lead wavefunctions  and the local wavefunction of the zero mode at the point of contact of the lead to the NW given by $u_{\alpha}$=$\langle\Phi_{\alpha}|\Psi (z)\rangle$. Here, $z$ represents the position of the tunneling contact 
of the NW to the lead. Thus these coefficients, describing the coupling of the zero mode to the leads follow the same spatial profile as the profile of the localized wavefunction corresponding to the zero mode. Note that, we restrict ourselves to zero bias conductance and hence all other modes which have a finite gap do not contribute to the conductance.  

In Fig.~\ref{fig6}, we have illlustrated the differential conductance at zero bias given by $G=dI/dV\lvert_{V=0}$ measured between the left and right lead as a function of the position of the left lead ($z_{-}=z<0$) and the right lead ($z_{+}=z>0$).  Here, $G_{0}=e^2/h$ expressed is the unit of quantized conductance. Note that, the zero energy conductance peak due to the Jackiw-Rebbi zero mode is widely spread over distance as depicted in Fig.~\ref{fig6}(a). On the other hand, Fig.~\ref{fig6}(b) illustrates the fact that 
upon increasing the value of angular momentum $l$, the $E=0$ conductance spectra gradually becomes localized over space. Therefore, to obtain the maximum differential conductance of magnitude 
$e^2/h$, via the single zero mode in our TI NWs set-up, one needs to place the two leads (see Fig.~\ref{fig1}) very sensitively on the two sides of the junction. In this way, the leads will locally 
attain the dominant contribution appearing from the single zero-mode only and the differential conductance reaches a maximum value of $e^2/h$. 
Moreover, the conductance signature will not be affected by the other mode present at $\pm \infty$. In Figs.~\ref{fig6}(a)-(b), the black 
dotted line is a guide to eye to denote the line of maximum differential conductance $e^2/h$ in the $z_{+}$-$z_{-}$ plane.
As the zero mode wavefunction is itself not symmetric about the junction, hence the quantized conductance peak is also not symmetric when plotted in the $z_{+}$-$z_{-}$ plane.  
Such a zero bias peak in the differential conductance can be a possible transport signature of the Jackiw-Rebbi zero mode appearing at the junction of two 3D TI NWs.

{\it{Summary and Conclusions:-}}
To summarize, in this article, we investigate the emergence of Jackiw-Rebbi zero modes in junction of 3D TI NWs. In the continuum limit, we first show that the surface states of  single cylindrical TI NW exhibits finite size gap in the spectrum whose sign depends on the algular momentum quantum number $l$ due to the confinement of electrons to the surface of the cylinder. Then applying a magnetic flux through the core of the cylinder, one can show that this gap, which represents a mass gap in 1D Dirac spectrum, can be tuned to zero when the Berry phase accumulated by the spins rotating on the surface of the cylinder exactly cancels the Aharonov-Bohm phase due to the flux. To realize JRMs, we consider a junction of two 3D TI NWs with different radii and demonstrate that suitable tuning of the magnetic flux can cause mass gap inversion across the junction for special values of $l$. Hence, localized Jackiw-Rebbi zero modes appear at the junction.  The degree of localization of the modes depends on the value of angular momentum quantum number $l$ and hence on the magnitude of the mass gap. 

We also establish the robustness of such modes against the rotation of the spin accross the junction due to variation of the spin-orbit coupling accross the junction. Finally, we explore the zero energy ($E=0$) 
differential conductance spectra through the zero mode by locally attaching it to two normal metallic leads. Note that, in our set-up, the leads are not attached to the two TI NWs at their ends.
Rather, the differential conductance attains a maximum value $e^2/h$, when the leads are appropriately placed about the junction of two TI NWs attaining dominant contribution from the single zero mode only
without being unaffected by the other mode present at $\pm \infty$. Such localized zero bias conductance can be a possible transport signal of the Jackiw-Rebbi zero mode proposed in our TI NW geometry. 

As far as validity of our results is concerned, we would like to point out that in general the continuity of the full 3D wave-function (which includes the radial coordinates $r$ of the NW) should be considered. Nevertheless, in our analysis, the Hamiltonian (Eq.(\ref{eq2})) is valid in the low energy limit where the surface states are well-defined and separated from bulk states by a gap. In each of the NWs the radius is fixed and effect due to their presence can be incorporated as a presence of a barrier at the junction. This can be understood as follows. In the above calculation we have considered a two band model to describe the surface states of TI NW which is a good description as long as we restrict our analysis to a single TI NW. However, for describing a junction between two different TI NWs we need to be more careful. The primary point lays in the fact that description of TI materials like $\rm Bi_{2}Se_{3}$ at low energies require a four band model~\cite{zhang2009topological}. As far as the surface states of these TIs are concerned, their wavefunction can be described as a direct product of two SU(2) degrees of freedom (totalling to four degrees of freedom) of which one is a dispersing (momentum dependent) SU(2) degree of freedom which is described by the two band model (see Eq.(\ref{eq2})) used above and a non-dispersing frozen (no momentum dependence) SU(2) degrees of freedom~\cite{PhysRevB.86.081303} which represents the existence of degrees of freedom of electron transverse to the surface. In the present problem of TI NW, the non-dispersing degrees of freedom would correspond to the dependence of the surface state wave function on radial coordinate of NW. It has been shown in Ref.~[\onlinecite{Roy_2016}] that the effect of such non-dispersing degree of freedom can lead to formation of sharp barrier for electron tunnelling at the junction between the two TIs. In our set-up, the two nanowires are different from each other due to this frozen degree of freedom owing to different radius $R_{1}$ and $R_{2}$. At the junction ($z=0$) of these two cylinders, the radial coordinate changes discontinuously from $R_{1}$ to $R_{2}$ as a function of $z$. Therefore, one can think of this situation as, at the junction of two distinct TI NWs with radius $R_{1}$ and $R_{2}$, the tunnelling of electron form one to the other depends on the overlap of wavefunction corresponding to the two different frozen degrees of freedom which in general would be different form unity leading to an effective potential barrier at the junction and the height of that potential would depend on the mismatch of the radial part of the wavefucntion. 
The presence of such constant potential barrier (not explicitly considered in our analysis) will introduce enhanced back-scattering of electrons from the junction which has been incorporated 
in our analysis via the presence of finite reflection amplitude while computing the transport signature of the Jackiw-Rebbi mode employing $\rm Weidenm\ddot{u}ller$ formula (Eqs.(\ref{eqr}-\ref{eqt})). 
Also, for moderate barrier height, the Jackiew-Rebbi mode would be more sharply localized at the junction and the maximum of the zero bias conductance will become less than $e^2/h$. 
Other than that, our results will remain qualitatively unchanged.

Earlier theoretical set-ups, based on 1D NWs, require either periodic modulation of the chemical potential~\cite{gangadharaiah2012localized} or simultaneous presence of strong Rashba SOI along with uniform and spatially periodic magnetic field~\cite{klinovaja2012transition, rainis2014transport} in order to realize JRMs in them. It may be a challenging task to tune all these ingredients together in an experimental set-up. On the other hand, in our 3D TI NW junction, the basic ingredient to tune the band inversion, accross the junction, is the uniform external magnetic field.

Thus, from the view point of experimental implementation of the proposal, our set-up can be more advantageous over the other set-ups as far as realization of JRMs is concerned. 

As far as practical realization of our geometry is concerned, very recently TI NW has been experimentally fabricated in 3D TI crystals ($\rm Bi_{1.33}Sb_{0.67}Se_{3}$) by mechanical exfoliation~\cite{cho2015aharonov}. The magnetic flux through the NW can be tuned by applying an external magnetic field~\cite{cho2015aharonov}. Hence, it may be experimentally possible to design  a junction of two TI NWs with radii $R_{1}\approx50~\rm nm$ and $R_{2}\approx150~\rm nm$~\cite{cho2015aharonov}. To realize the zero modes, the chemical potential and the magnetic flux
inside the two NWs can be tuned to $\mu_{1}\approx \rm 10~meV$, $\mu_{2}\approx \rm 3.5~meV$ and $\Phi_{1}\approx \rm 8 \times 10^{-15}~Wb$, $\Phi_{2}\approx \rm 7 \times 10^{-14}~Wb$ respectively 
for a magnetic field of $B\approx \rm 1~T$. The transport signature of the zero modes should be robust in presence of scalar disorder as far as the disorder scale is smaller than the band gap of the cylindrical 
TI NW.

{\it{Acknowledgements:-}} SD acknowledges the warm hospitality of IOP Bhubaneswar where this work was initiated. SJ and AS acknowledge the warm hospitality of IISER kolkata 
where part of this work was performed.



\bibliography{bibfile}{}

\begin{thebibliography}{46}%
\makeatletter
\providecommand \@ifxundefined [1]{%
 \@ifx{#1\undefined}
}%
\providecommand \@ifnum [1]{%
 \ifnum #1\expandafter \@firstoftwo
 \else \expandafter \@secondoftwo
 \fi
}%
\providecommand \@ifx [1]{%
 \ifx #1\expandafter \@firstoftwo
 \else \expandafter \@secondoftwo
 \fi
}%
\providecommand \natexlab [1]{#1}%
\providecommand \enquote  [1]{``#1''}%
\providecommand \bibnamefont  [1]{#1}%
\providecommand \bibfnamefont [1]{#1}%
\providecommand \citenamefont [1]{#1}%
\providecommand \href@noop [0]{\@secondoftwo}%
\providecommand \href [0]{\begingroup \@sanitize@url \@href}%
\providecommand \@href[1]{\@@startlink{#1}\@@href}%
\providecommand \@@href[1]{\endgroup#1\@@endlink}%
\providecommand \@sanitize@url [0]{\catcode `\\12\catcode `\$12\catcode
  `\&12\catcode `\#12\catcode `\^12\catcode `\_12\catcode `\%12\relax}%
\providecommand \@@startlink[1]{}%
\providecommand \@@endlink[0]{}%
\providecommand \url  [0]{\begingroup\@sanitize@url \@url }%
\providecommand \@url [1]{\endgroup\@href {#1}{\urlprefix }}%
\providecommand \urlprefix  [0]{URL }%
\providecommand \Eprint [0]{\href }%
\providecommand \doibase [0]{http://dx.doi.org/}%
\providecommand \selectlanguage [0]{\@gobble}%
\providecommand \bibinfo  [0]{\@secondoftwo}%
\providecommand \bibfield  [0]{\@secondoftwo}%
\providecommand \translation [1]{[#1]}%
\providecommand \BibitemOpen [0]{}%
\providecommand \bibitemStop [0]{}%
\providecommand \bibitemNoStop [0]{.\EOS\space}%
\providecommand \EOS [0]{\spacefactor3000\relax}%
\providecommand \BibitemShut  [1]{\csname bibitem#1\endcsname}%
\let\auto@bib@innerbib\@empty
\bibitem [{\citenamefont {Nayak}\ \emph {et~al.}(2008)\citenamefont {Nayak},
  \citenamefont {Simon}, \citenamefont {Stern}, \citenamefont {Freedman},\ and\
  \citenamefont {Sarma}}]{nayak2008non}%
  \BibitemOpen
  \bibfield  {author} {\bibinfo {author} {\bibfnamefont {C.}~\bibnamefont
  {Nayak}}, \bibinfo {author} {\bibfnamefont {S.~H.}\ \bibnamefont {Simon}},
  \bibinfo {author} {\bibfnamefont {A.}~\bibnamefont {Stern}}, \bibinfo
  {author} {\bibfnamefont {M.}~\bibnamefont {Freedman}}, \ and\ \bibinfo
  {author} {\bibfnamefont {S.~D.}\ \bibnamefont {Sarma}},\ }\href@noop {}
  {\bibfield  {journal} {\bibinfo  {journal} {Rev. Mod. Phys.}\ }\textbf
  {\bibinfo {volume} {80}},\ \bibinfo {pages} {1083} (\bibinfo {year}
  {2008})}\BibitemShut {NoStop}%
\bibitem [{\citenamefont {Jackiw}\ and\ \citenamefont
  {Rebbi}(1976)}]{jackiw1976solitons}%
  \BibitemOpen
  \bibfield  {author} {\bibinfo {author} {\bibfnamefont {R.}~\bibnamefont
  {Jackiw}}\ and\ \bibinfo {author} {\bibfnamefont {C.}~\bibnamefont {Rebbi}},\
  }\href@noop {} {\bibfield  {journal} {\bibinfo  {journal} {Phys. Rev. D}\
  }\textbf {\bibinfo {volume} {13}},\ \bibinfo {pages} {3398} (\bibinfo {year}
  {1976})}\BibitemShut {NoStop}%
\bibitem [{\citenamefont {Rajaraman}\ and\ \citenamefont
  {Bell}(1982)}]{RajaramanBell1982}%
  \BibitemOpen
  \bibfield  {author} {\bibinfo {author} {\bibfnamefont {R.}~\bibnamefont
  {Rajaraman}}\ and\ \bibinfo {author} {\bibfnamefont {J.~S.}\ \bibnamefont
  {Bell}},\ }\href@noop {} {\bibfield  {journal} {\bibinfo  {journal} {Phys.
  Lett. B}\ }\textbf {\bibinfo {volume} {116}},\ \bibinfo {pages} {151}
  (\bibinfo {year} {1982})}\BibitemShut {NoStop}%
\bibitem [{\citenamefont {Rajaraman}()}]{Rajaraman2001}%
  \BibitemOpen
  \bibfield  {author} {\bibinfo {author} {\bibfnamefont {R.}~\bibnamefont
  {Rajaraman}},\ }\href@noop {} {\bibinfo  {journal} {arXiv:cond-mat/0103366
  [cond-mat.mes-hall]}\ }\BibitemShut {NoStop}%
\bibitem [{\citenamefont {Su}\ \emph {et~al.}(1979)\citenamefont {Su},
  \citenamefont {Schrieffer},\ and\ \citenamefont {Heeger}}]{su1980solitonPRL}%
  \BibitemOpen
\bibfield  {journal} {  }\bibfield  {author} {\bibinfo {author} {\bibfnamefont
  {W.-P.}\ \bibnamefont {Su}}, \bibinfo {author} {\bibfnamefont
  {J.}~\bibnamefont {Schrieffer}}, \ and\ \bibinfo {author} {\bibfnamefont
  {A.}~\bibnamefont {Heeger}},\ }\href@noop {} {\bibfield  {journal} {\bibinfo
  {journal} {Phys. Rev. Lett.}\ }\textbf {\bibinfo {volume} {42}},\ \bibinfo
  {pages} {1698} (\bibinfo {year} {1979})}\BibitemShut {NoStop}%
\bibitem [{\citenamefont {Su}\ \emph {et~al.}(1980)\citenamefont {Su},
  \citenamefont {Schrieffer},\ and\ \citenamefont {Heeger}}]{su1980soliton}%
  \BibitemOpen
  \bibfield  {author} {\bibinfo {author} {\bibfnamefont {W.-P.}\ \bibnamefont
  {Su}}, \bibinfo {author} {\bibfnamefont {J.}~\bibnamefont {Schrieffer}}, \
  and\ \bibinfo {author} {\bibfnamefont {A.}~\bibnamefont {Heeger}},\
  }\href@noop {} {\bibfield  {journal} {\bibinfo  {journal} {Phys. Rev. B}\
  }\textbf {\bibinfo {volume} {22}},\ \bibinfo {pages} {2099} (\bibinfo {year}
  {1980})}\BibitemShut {NoStop}%
\bibitem [{\citenamefont {Gangadharaiah}\ \emph {et~al.}(2012)\citenamefont
  {Gangadharaiah}, \citenamefont {Trifunovic},\ and\ \citenamefont
  {Loss}}]{gangadharaiah2012localized}%
  \BibitemOpen
  \bibfield  {author} {\bibinfo {author} {\bibfnamefont {S.}~\bibnamefont
  {Gangadharaiah}}, \bibinfo {author} {\bibfnamefont {L.}~\bibnamefont
  {Trifunovic}}, \ and\ \bibinfo {author} {\bibfnamefont {D.}~\bibnamefont
  {Loss}},\ }\href@noop {} {\bibfield  {journal} {\bibinfo  {journal} {Phys.
  Rev. Lett.}\ }\textbf {\bibinfo {volume} {108}},\ \bibinfo {pages} {136803}
  (\bibinfo {year} {2012})}\BibitemShut {NoStop}%
\bibitem [{\citenamefont {Park}\ \emph {et~al.}(2016)\citenamefont {Park},
  \citenamefont {Yang}, \citenamefont {Klinovaja}, \citenamefont {Stano},\ and\
  \citenamefont {Loss}}]{PhysRevB.94.075416}%
  \BibitemOpen
  \bibfield  {author} {\bibinfo {author} {\bibfnamefont {J.-H.}\ \bibnamefont
  {Park}}, \bibinfo {author} {\bibfnamefont {G.}~\bibnamefont {Yang}}, \bibinfo
  {author} {\bibfnamefont {J.}~\bibnamefont {Klinovaja}}, \bibinfo {author}
  {\bibfnamefont {P.}~\bibnamefont {Stano}}, \ and\ \bibinfo {author}
  {\bibfnamefont {D.}~\bibnamefont {Loss}},\ }\href@noop {} {\bibfield
  {journal} {\bibinfo  {journal} {Phys. Rev. B}\ }\textbf {\bibinfo {volume}
  {94}},\ \bibinfo {pages} {075416} (\bibinfo {year} {2016})}\BibitemShut
  {NoStop}%
\bibitem [{\citenamefont {V\"ayrynen}\ and\ \citenamefont
  {Ojanen}(2011)}]{PhysRevLett.107.166804}%
  \BibitemOpen
  \bibfield  {author} {\bibinfo {author} {\bibfnamefont {J.~I.}\ \bibnamefont
  {V\"ayrynen}}\ and\ \bibinfo {author} {\bibfnamefont {T.}~\bibnamefont
  {Ojanen}},\ }\href@noop {} {\bibfield  {journal} {\bibinfo  {journal} {Phys.
  Rev. Lett.}\ }\textbf {\bibinfo {volume} {107}},\ \bibinfo {pages} {166804}
  (\bibinfo {year} {2011})}\BibitemShut {NoStop}%
\bibitem [{\citenamefont {Qi}\ \emph {et~al.}(2008)\citenamefont {Qi},
  \citenamefont {Hughes},\ and\ \citenamefont {Zhang}}]{qi2008fractional}%
  \BibitemOpen
  \bibfield  {author} {\bibinfo {author} {\bibfnamefont {X.-L.}\ \bibnamefont
  {Qi}}, \bibinfo {author} {\bibfnamefont {T.~L.}\ \bibnamefont {Hughes}}, \
  and\ \bibinfo {author} {\bibfnamefont {S.-C.}\ \bibnamefont {Zhang}},\
  }\href@noop {} {\bibfield  {journal} {\bibinfo  {journal} {Nat. Phys.}\
  }\textbf {\bibinfo {volume} {4}},\ \bibinfo {pages} {273} (\bibinfo {year}
  {2008})}\BibitemShut {NoStop}%
\bibitem [{\citenamefont {Fleckenstein}\ \emph {et~al.}(2016)\citenamefont
  {Fleckenstein}, \citenamefont {Traverso~Ziani},\ and\ \citenamefont
  {Trauzettel}}]{PhysRevB.94.241406}%
  \BibitemOpen
  \bibfield  {author} {\bibinfo {author} {\bibfnamefont {C.}~\bibnamefont
  {Fleckenstein}}, \bibinfo {author} {\bibfnamefont {N.}~\bibnamefont
  {Traverso~Ziani}}, \ and\ \bibinfo {author} {\bibfnamefont {B.}~\bibnamefont
  {Trauzettel}},\ }\href@noop {} {\bibfield  {journal} {\bibinfo  {journal}
  {Phys. Rev. B}\ }\textbf {\bibinfo {volume} {94}},\ \bibinfo {pages} {241406}
  (\bibinfo {year} {2016})}\BibitemShut {NoStop}%
\bibitem [{\citenamefont {Klinovaja}\ \emph {et~al.}(2012)\citenamefont
  {Klinovaja}, \citenamefont {Stano},\ and\ \citenamefont
  {Loss}}]{klinovaja2012transition}%
  \BibitemOpen
  \bibfield  {author} {\bibinfo {author} {\bibfnamefont {J.}~\bibnamefont
  {Klinovaja}}, \bibinfo {author} {\bibfnamefont {P.}~\bibnamefont {Stano}}, \
  and\ \bibinfo {author} {\bibfnamefont {D.}~\bibnamefont {Loss}},\ }\href@noop
  {} {\bibfield  {journal} {\bibinfo  {journal} {Phys. Rev. Lett.}\ }\textbf
  {\bibinfo {volume} {109}},\ \bibinfo {pages} {236801} (\bibinfo {year}
  {2012})}\BibitemShut {NoStop}%
\bibitem [{\citenamefont {Rainis}\ \emph {et~al.}(2014)\citenamefont {Rainis},
  \citenamefont {Saha}, \citenamefont {Klinovaja}, \citenamefont {Trifunovic},\
  and\ \citenamefont {Loss}}]{rainis2014transport}%
  \BibitemOpen
  \bibfield  {author} {\bibinfo {author} {\bibfnamefont {D.}~\bibnamefont
  {Rainis}}, \bibinfo {author} {\bibfnamefont {A.}~\bibnamefont {Saha}},
  \bibinfo {author} {\bibfnamefont {J.}~\bibnamefont {Klinovaja}}, \bibinfo
  {author} {\bibfnamefont {L.}~\bibnamefont {Trifunovic}}, \ and\ \bibinfo
  {author} {\bibfnamefont {D.}~\bibnamefont {Loss}},\ }\href@noop {} {\bibfield
   {journal} {\bibinfo  {journal} {Phys. Rev. Lett.}\ }\textbf {\bibinfo
  {volume} {112}},\ \bibinfo {pages} {196803} (\bibinfo {year}
  {2014})}\BibitemShut {NoStop}%
\bibitem [{\citenamefont {Klinovaja}\ and\ \citenamefont
  {Loss}(2015)}]{JelenaLoss}%
  \BibitemOpen
  \bibfield  {author} {\bibinfo {author} {\bibfnamefont {J.}~\bibnamefont
  {Klinovaja}}\ and\ \bibinfo {author} {\bibfnamefont {D.}~\bibnamefont
  {Loss}},\ }\href@noop {} {\bibfield  {journal} {\bibinfo  {journal} {Eur.
  Phys. J. B}\ }\textbf {\bibinfo {volume} {88}},\ \bibinfo {pages} {62}
  (\bibinfo {year} {2015})}\BibitemShut {NoStop}%
\bibitem [{\citenamefont {Klinovaja}\ and\ \citenamefont
  {Loss}(2013)}]{PhysRevLett.110.126402}%
  \BibitemOpen
  \bibfield  {author} {\bibinfo {author} {\bibfnamefont {J.}~\bibnamefont
  {Klinovaja}}\ and\ \bibinfo {author} {\bibfnamefont {D.}~\bibnamefont
  {Loss}},\ }\href@noop {} {\bibfield  {journal} {\bibinfo  {journal} {Phys.
  Rev. Lett.}\ }\textbf {\bibinfo {volume} {110}},\ \bibinfo {pages} {126402}
  (\bibinfo {year} {2013})}\BibitemShut {NoStop}%
\bibitem [{\citenamefont {Saha}\ \emph {et~al.}(2014)\citenamefont {Saha},
  \citenamefont {Rainis}, \citenamefont {Tiwari},\ and\ \citenamefont
  {Loss}}]{saha2014quantum}%
  \BibitemOpen
  \bibfield  {author} {\bibinfo {author} {\bibfnamefont {A.}~\bibnamefont
  {Saha}}, \bibinfo {author} {\bibfnamefont {D.}~\bibnamefont {Rainis}},
  \bibinfo {author} {\bibfnamefont {R.~P.}\ \bibnamefont {Tiwari}}, \ and\
  \bibinfo {author} {\bibfnamefont {D.}~\bibnamefont {Loss}},\ }\href@noop {}
  {\bibfield  {journal} {\bibinfo  {journal} {Phys. Rev. B}\ }\textbf {\bibinfo
  {volume} {90}},\ \bibinfo {pages} {035422} (\bibinfo {year}
  {2014})}\BibitemShut {NoStop}%
\bibitem [{\citenamefont {Kane}\ and\ \citenamefont
  {Mele}(2005)}]{kane2005quantum}%
  \BibitemOpen
  \bibfield  {author} {\bibinfo {author} {\bibfnamefont {C.~L.}\ \bibnamefont
  {Kane}}\ and\ \bibinfo {author} {\bibfnamefont {E.~J.}\ \bibnamefont
  {Mele}},\ }\href@noop {} {\bibfield  {journal} {\bibinfo  {journal} {Phys.
  Rev. Lett.}\ }\textbf {\bibinfo {volume} {95}},\ \bibinfo {pages} {226801}
  (\bibinfo {year} {2005})}\BibitemShut {NoStop}%
\bibitem [{\citenamefont {Bernevig}\ \emph {et~al.}(2006)\citenamefont
  {Bernevig}, \citenamefont {Hughes},\ and\ \citenamefont
  {Zhang}}]{bernevig2006quantum}%
  \BibitemOpen
  \bibfield  {author} {\bibinfo {author} {\bibfnamefont {B.~A.}\ \bibnamefont
  {Bernevig}}, \bibinfo {author} {\bibfnamefont {T.~L.}\ \bibnamefont
  {Hughes}}, \ and\ \bibinfo {author} {\bibfnamefont {S.-C.}\ \bibnamefont
  {Zhang}},\ }\href@noop {} {\bibfield  {journal} {\bibinfo  {journal}
  {Science}\ }\textbf {\bibinfo {volume} {314}},\ \bibinfo {pages} {1757}
  (\bibinfo {year} {2006})}\BibitemShut {NoStop}%
\bibitem [{\citenamefont {Moore}(2010)}]{moore2010birth}%
  \BibitemOpen
  \bibfield  {author} {\bibinfo {author} {\bibfnamefont {J.~E.}\ \bibnamefont
  {Moore}},\ }\href@noop {} {\bibfield  {journal} {\bibinfo  {journal}
  {Nature}\ }\textbf {\bibinfo {volume} {464}},\ \bibinfo {pages} {194}
  (\bibinfo {year} {2010})}\BibitemShut {NoStop}%
\bibitem [{\citenamefont {Hasan}\ and\ \citenamefont
  {Kane}(2010)}]{hasan2010colloquium}%
  \BibitemOpen
  \bibfield  {author} {\bibinfo {author} {\bibfnamefont {M.~Z.}\ \bibnamefont
  {Hasan}}\ and\ \bibinfo {author} {\bibfnamefont {C.~L.}\ \bibnamefont
  {Kane}},\ }\href@noop {} {\bibfield  {journal} {\bibinfo  {journal} {Rev.
  Mod. Phys.}\ }\textbf {\bibinfo {volume} {82}},\ \bibinfo {pages} {3045}
  (\bibinfo {year} {2010})}\BibitemShut {NoStop}%
\bibitem [{\citenamefont {Qi}\ and\ \citenamefont
  {Zhang}(2011)}]{qi2011topological}%
  \BibitemOpen
  \bibfield  {author} {\bibinfo {author} {\bibfnamefont {X.-L.}\ \bibnamefont
  {Qi}}\ and\ \bibinfo {author} {\bibfnamefont {S.-C.}\ \bibnamefont {Zhang}},\
  }\href@noop {} {\bibfield  {journal} {\bibinfo  {journal} {Rev. Mod. Phys.}\
  }\textbf {\bibinfo {volume} {83}},\ \bibinfo {pages} {1057} (\bibinfo {year}
  {2011})}\BibitemShut {NoStop}%
\bibitem [{\citenamefont {K{\"o}nig}\ \emph {et~al.}(2007)\citenamefont
  {K{\"o}nig}, \citenamefont {Wiedmann}, \citenamefont {Br{\"u}ne},
  \citenamefont {Roth}, \citenamefont {Buhmann}, \citenamefont {Molenkamp},
  \citenamefont {Qi},\ and\ \citenamefont {Zhang}}]{konig2007quantum}%
  \BibitemOpen
  \bibfield  {author} {\bibinfo {author} {\bibfnamefont {M.}~\bibnamefont
  {K{\"o}nig}}, \bibinfo {author} {\bibfnamefont {S.}~\bibnamefont {Wiedmann}},
  \bibinfo {author} {\bibfnamefont {C.}~\bibnamefont {Br{\"u}ne}}, \bibinfo
  {author} {\bibfnamefont {A.}~\bibnamefont {Roth}}, \bibinfo {author}
  {\bibfnamefont {H.}~\bibnamefont {Buhmann}}, \bibinfo {author} {\bibfnamefont
  {L.~W.}\ \bibnamefont {Molenkamp}}, \bibinfo {author} {\bibfnamefont {X.-L.}\
  \bibnamefont {Qi}}, \ and\ \bibinfo {author} {\bibfnamefont {S.-C.}\
  \bibnamefont {Zhang}},\ }\href@noop {} {\bibfield  {journal} {\bibinfo
  {journal} {Science}\ }\textbf {\bibinfo {volume} {318}},\ \bibinfo {pages}
  {766} (\bibinfo {year} {2007})}\BibitemShut {NoStop}%
\bibitem [{\citenamefont {Culcer}(2012)}]{culcer2012transport}%
  \BibitemOpen
  \bibfield  {author} {\bibinfo {author} {\bibfnamefont {D.}~\bibnamefont
  {Culcer}},\ }\href@noop {} {\bibfield  {journal} {\bibinfo  {journal}
  {Physica E: Low-dimensional Systems and Nanostructures}\ }\textbf {\bibinfo
  {volume} {44}},\ \bibinfo {pages} {860} (\bibinfo {year} {2012})}\BibitemShut
  {NoStop}%
\bibitem [{\citenamefont {Xia}\ \emph {et~al.}(2009)\citenamefont {Xia},
  \citenamefont {Qian}, \citenamefont {Hsieh}, \citenamefont {Wray},
  \citenamefont {Pal}, \citenamefont {Lin}, \citenamefont {Bansil},
  \citenamefont {Grauer}, \citenamefont {Hor}, \citenamefont {Cava} \emph
  {et~al.}}]{xia2009observation}%
  \BibitemOpen
  \bibfield  {author} {\bibinfo {author} {\bibfnamefont {Y.}~\bibnamefont
  {Xia}}, \bibinfo {author} {\bibfnamefont {D.}~\bibnamefont {Qian}}, \bibinfo
  {author} {\bibfnamefont {D.}~\bibnamefont {Hsieh}}, \bibinfo {author}
  {\bibfnamefont {L.}~\bibnamefont {Wray}}, \bibinfo {author} {\bibfnamefont
  {A.}~\bibnamefont {Pal}}, \bibinfo {author} {\bibfnamefont {H.}~\bibnamefont
  {Lin}}, \bibinfo {author} {\bibfnamefont {A.}~\bibnamefont {Bansil}},
  \bibinfo {author} {\bibfnamefont {D.}~\bibnamefont {Grauer}}, \bibinfo
  {author} {\bibfnamefont {Y.~S.}\ \bibnamefont {Hor}}, \bibinfo {author}
  {\bibfnamefont {R.~J.}\ \bibnamefont {Cava}},  \emph {et~al.},\ }\href@noop
  {} {\bibfield  {journal} {\bibinfo  {journal} {Nat. Phys.}\ }\textbf
  {\bibinfo {volume} {5}},\ \bibinfo {pages} {398} (\bibinfo {year}
  {2009})}\BibitemShut {NoStop}%
\bibitem [{\citenamefont {Fu}\ and\ \citenamefont
  {Kane}(2008)}]{fu2008superconducting}%
  \BibitemOpen
  \bibfield  {author} {\bibinfo {author} {\bibfnamefont {L.}~\bibnamefont
  {Fu}}\ and\ \bibinfo {author} {\bibfnamefont {C.~L.}\ \bibnamefont {Kane}},\
  }\href@noop {} {\bibfield  {journal} {\bibinfo  {journal} {Phys. Rev. Lett.}\
  }\textbf {\bibinfo {volume} {100}},\ \bibinfo {pages} {096407} (\bibinfo
  {year} {2008})}\BibitemShut {NoStop}%
\bibitem [{\citenamefont {Alicea}(2012)}]{alicea2012new}%
  \BibitemOpen
  \bibfield  {author} {\bibinfo {author} {\bibfnamefont {J.}~\bibnamefont
  {Alicea}},\ }\href@noop {} {\bibfield  {journal} {\bibinfo  {journal} {Rep.
  Prog. Phys.}\ }\textbf {\bibinfo {volume} {75}},\ \bibinfo {pages} {076501}
  (\bibinfo {year} {2012})}\BibitemShut {NoStop}%
\bibitem [{\citenamefont {Cook}\ and\ \citenamefont
  {Franz}(2011)}]{cook2011majorana}%
  \BibitemOpen
  \bibfield  {author} {\bibinfo {author} {\bibfnamefont {A.}~\bibnamefont
  {Cook}}\ and\ \bibinfo {author} {\bibfnamefont {M.}~\bibnamefont {Franz}},\
  }\href@noop {} {\bibfield  {journal} {\bibinfo  {journal} {Phys. Rev. B}\
  }\textbf {\bibinfo {volume} {84}},\ \bibinfo {pages} {201105} (\bibinfo
  {year} {2011})}\BibitemShut {NoStop}%
\bibitem [{\citenamefont {de~Juan}\ \emph {et~al.}(2014)\citenamefont
  {de~Juan}, \citenamefont {Ilan},\ and\ \citenamefont
  {Bardarson}}]{roni2014PRL}%
  \BibitemOpen
  \bibfield  {author} {\bibinfo {author} {\bibfnamefont {F.}~\bibnamefont
  {de~Juan}}, \bibinfo {author} {\bibfnamefont {R.}~\bibnamefont {Ilan}}, \
  and\ \bibinfo {author} {\bibfnamefont {J.~H.}\ \bibnamefont {Bardarson}},\
  }\href@noop {} {\bibfield  {journal} {\bibinfo  {journal} {Phys. Rev. Lett.}\
  }\textbf {\bibinfo {volume} {113}},\ \bibinfo {pages} {107003} (\bibinfo
  {year} {2014})}\BibitemShut {NoStop}%
\bibitem [{\citenamefont {Huang}\ and\ \citenamefont
  {Xu}(2017)}]{GYHuangPRB2017}%
  \BibitemOpen
  \bibfield  {author} {\bibinfo {author} {\bibfnamefont {G.~Y.}\ \bibnamefont
  {Huang}}\ and\ \bibinfo {author} {\bibfnamefont {H.~Q.}\ \bibnamefont {Xu}},\
  }\href@noop {} {\bibfield  {journal} {\bibinfo  {journal} {Phys. Rev. B}\
  }\textbf {\bibinfo {volume} {95}},\ \bibinfo {pages} {155420} (\bibinfo
  {year} {2017})}\BibitemShut {NoStop}%
\bibitem [{\citenamefont {de~Juan}\ \emph {et~al.}()\citenamefont {de~Juan},
  \citenamefont {Bardarson},\ and\ \citenamefont {Ilan}}]{de2018conditions}%
  \BibitemOpen
  \bibfield  {author} {\bibinfo {author} {\bibfnamefont {F.}~\bibnamefont
  {de~Juan}}, \bibinfo {author} {\bibfnamefont {J.~H.}\ \bibnamefont
  {Bardarson}}, \ and\ \bibinfo {author} {\bibfnamefont {R.}~\bibnamefont
  {Ilan}},\ }\href@noop {} {\bibinfo  {journal} {arXiv:1810.09576
  [cond-mat.mess-hall]}\ }\BibitemShut {NoStop}%
\bibitem [{\citenamefont {Ran}\ \emph {et~al.}(2008)\citenamefont {Ran},
  \citenamefont {Vishwanath},\ and\ \citenamefont {Lee}}]{ran2008spin}%
  \BibitemOpen
\bibfield  {journal} {  }\bibfield  {author} {\bibinfo {author} {\bibfnamefont
  {Y.}~\bibnamefont {Ran}}, \bibinfo {author} {\bibfnamefont {A.}~\bibnamefont
  {Vishwanath}}, \ and\ \bibinfo {author} {\bibfnamefont {D.-H.}\ \bibnamefont
  {Lee}},\ }\href@noop {} {\bibfield  {journal} {\bibinfo  {journal} {Phys.
  Rev. Lett.}\ }\textbf {\bibinfo {volume} {101}},\ \bibinfo {pages} {086801}
  (\bibinfo {year} {2008})}\BibitemShut {NoStop}%
\bibitem [{\citenamefont {Rosenberg}\ \emph {et~al.}(2010)\citenamefont
  {Rosenberg}, \citenamefont {Guo},\ and\ \citenamefont
  {Franz}}]{rosenberg2010wormhole}%
  \BibitemOpen
  \bibfield  {author} {\bibinfo {author} {\bibfnamefont {G.}~\bibnamefont
  {Rosenberg}}, \bibinfo {author} {\bibfnamefont {H.-M.}\ \bibnamefont {Guo}},
  \ and\ \bibinfo {author} {\bibfnamefont {M.}~\bibnamefont {Franz}},\
  }\href@noop {} {\bibfield  {journal} {\bibinfo  {journal} {Phys. Rev. B}\
  }\textbf {\bibinfo {volume} {82}},\ \bibinfo {pages} {041104} (\bibinfo
  {year} {2010})}\BibitemShut {NoStop}%
\bibitem [{\citenamefont {Juri\ifmmode \check{c}\else
  \v{c}\fi{}i\ifmmode~\acute{c}\else \'{c}\fi{}}\ \emph
  {et~al.}(2012)\citenamefont {Juri\ifmmode \check{c}\else
  \v{c}\fi{}i\ifmmode~\acute{c}\else \'{c}\fi{}}, \citenamefont {Mesaros},
  \citenamefont {Slager},\ and\ \citenamefont
  {Zaanen}}]{PhysRevLett.108.106403}%
  \BibitemOpen
  \bibfield  {author} {\bibinfo {author} {\bibfnamefont {V.}~\bibnamefont
  {Juri\ifmmode \check{c}\else \v{c}\fi{}i\ifmmode~\acute{c}\else \'{c}\fi{}}},
  \bibinfo {author} {\bibfnamefont {A.}~\bibnamefont {Mesaros}}, \bibinfo
  {author} {\bibfnamefont {R.-J.}\ \bibnamefont {Slager}}, \ and\ \bibinfo
  {author} {\bibfnamefont {J.}~\bibnamefont {Zaanen}},\ }\href@noop {}
  {\bibfield  {journal} {\bibinfo  {journal} {Phys. Rev. Lett.}\ }\textbf
  {\bibinfo {volume} {108}},\ \bibinfo {pages} {106403} (\bibinfo {year}
  {2012})}\BibitemShut {NoStop}%
\bibitem [{\citenamefont {Mesaros}\ \emph {et~al.}(2013)\citenamefont
  {Mesaros}, \citenamefont {Slager}, \citenamefont {Zaanen},\ and\
  \citenamefont {Juri\ifmmode \check{c}\else \v{c}\fi{}i\ifmmode~\acute{c}\else
  \'{c}\fi{}}}]{MESAROS2013977}%
  \BibitemOpen
  \bibfield  {author} {\bibinfo {author} {\bibfnamefont {A.}~\bibnamefont
  {Mesaros}}, \bibinfo {author} {\bibfnamefont {R.-J.}\ \bibnamefont {Slager}},
  \bibinfo {author} {\bibfnamefont {J.}~\bibnamefont {Zaanen}}, \ and\ \bibinfo
  {author} {\bibfnamefont {V.}~\bibnamefont {Juri\ifmmode \check{c}\else
  \v{c}\fi{}i\ifmmode~\acute{c}\else \'{c}\fi{}}},\ }\href@noop {} {\bibfield
  {journal} {\bibinfo  {journal} {Nuclear Physics B}\ }\textbf {\bibinfo
  {volume} {867}},\ \bibinfo {pages} {977} (\bibinfo {year}
  {2013})}\BibitemShut {NoStop}%
\bibitem [{\citenamefont {Ilan}\ \emph {et~al.}(2014)\citenamefont {Ilan},
  \citenamefont {Bardarson}, \citenamefont {Sim},\ and\ \citenamefont
  {Moore}}]{RoniIlanNJP2014}%
  \BibitemOpen
  \bibfield  {author} {\bibinfo {author} {\bibfnamefont {R.}~\bibnamefont
  {Ilan}}, \bibinfo {author} {\bibfnamefont {J.~H.}\ \bibnamefont {Bardarson}},
  \bibinfo {author} {\bibfnamefont {H.-S.}\ \bibnamefont {Sim}}, \ and\
  \bibinfo {author} {\bibfnamefont {J.~E.}\ \bibnamefont {Moore}},\ }\href@noop
  {} {\bibfield  {journal} {\bibinfo  {journal} {New J. Phys.}\ }\textbf
  {\bibinfo {volume} {16}},\ \bibinfo {pages} {053007} (\bibinfo {year}
  {2014})}\BibitemShut {NoStop}%
\bibitem [{\citenamefont {Moors}\ \emph {et~al.}(2018)\citenamefont {Moors},
  \citenamefont {Sch{\"u}ffelgen}, \citenamefont {Rosenbach}, \citenamefont
  {Schmitt}, \citenamefont {Sch{\"a}pers},\ and\ \citenamefont
  {Schmidt}}]{moors2018magnetotransport}%
  \BibitemOpen
  \bibfield  {author} {\bibinfo {author} {\bibfnamefont {K.}~\bibnamefont
  {Moors}}, \bibinfo {author} {\bibfnamefont {P.}~\bibnamefont
  {Sch{\"u}ffelgen}}, \bibinfo {author} {\bibfnamefont {D.}~\bibnamefont
  {Rosenbach}}, \bibinfo {author} {\bibfnamefont {T.}~\bibnamefont {Schmitt}},
  \bibinfo {author} {\bibfnamefont {T.}~\bibnamefont {Sch{\"a}pers}}, \ and\
  \bibinfo {author} {\bibfnamefont {T.~L.}\ \bibnamefont {Schmidt}},\
  }\href@noop {} {\bibfield  {journal} {\bibinfo  {journal} {Phys. Rev. B}\
  }\textbf {\bibinfo {volume} {97}},\ \bibinfo {pages} {245429} (\bibinfo
  {year} {2018})}\BibitemShut {NoStop}%
\bibitem [{\citenamefont {Erlingsson}\ \emph {et~al.}(2017)\citenamefont
  {Erlingsson}, \citenamefont {Manolescu}, \citenamefont {Nemnes},
  \citenamefont {Bardarson},\ and\ \citenamefont
  {Sanchez}}]{thermoelectricTINW2017}%
  \BibitemOpen
  \bibfield  {author} {\bibinfo {author} {\bibfnamefont {S.~I.}\ \bibnamefont
  {Erlingsson}}, \bibinfo {author} {\bibfnamefont {A.}~\bibnamefont
  {Manolescu}}, \bibinfo {author} {\bibfnamefont {G.~A.}\ \bibnamefont
  {Nemnes}}, \bibinfo {author} {\bibfnamefont {J.~H.}\ \bibnamefont
  {Bardarson}}, \ and\ \bibinfo {author} {\bibfnamefont {D.}~\bibnamefont
  {Sanchez}},\ }\href@noop {} {\bibfield  {journal} {\bibinfo  {journal} {Phys.
  Rev. Lett.}\ }\textbf {\bibinfo {volume} {119}},\ \bibinfo {pages} {036804}
  (\bibinfo {year} {2017})}\BibitemShut {NoStop}%
\bibitem [{\citenamefont {Cho}\ \emph {et~al.}(2015)\citenamefont {Cho},
  \citenamefont {Dellabetta}, \citenamefont {Zhong}, \citenamefont
  {Schneeloch}, \citenamefont {Liu}, \citenamefont {Gu}, \citenamefont
  {Gilbert},\ and\ \citenamefont {Mason}}]{cho2015aharonov}%
  \BibitemOpen
  \bibfield  {author} {\bibinfo {author} {\bibfnamefont {S.}~\bibnamefont
  {Cho}}, \bibinfo {author} {\bibfnamefont {B.}~\bibnamefont {Dellabetta}},
  \bibinfo {author} {\bibfnamefont {R.}~\bibnamefont {Zhong}}, \bibinfo
  {author} {\bibfnamefont {J.}~\bibnamefont {Schneeloch}}, \bibinfo {author}
  {\bibfnamefont {T.}~\bibnamefont {Liu}}, \bibinfo {author} {\bibfnamefont
  {G.}~\bibnamefont {Gu}}, \bibinfo {author} {\bibfnamefont {M.~J.}\
  \bibnamefont {Gilbert}}, \ and\ \bibinfo {author} {\bibfnamefont
  {N.}~\bibnamefont {Mason}},\ }\href@noop {} {\bibfield  {journal} {\bibinfo
  {journal} {Nature Communications}\ }\textbf {\bibinfo {volume} {6}},\
  \bibinfo {pages} {7634} (\bibinfo {year} {2015})}\BibitemShut {NoStop}%
\bibitem [{\citenamefont {Jauregui}\ \emph {et~al.}(2018)\citenamefont
  {Jauregui}, \citenamefont {Kayyalha}, \citenamefont {Kazakov}, \citenamefont
  {Miotkowski}, \citenamefont {Rokhinson},\ and\ \citenamefont
  {Chen}}]{APL2018}%
  \BibitemOpen
  \bibfield  {author} {\bibinfo {author} {\bibfnamefont {L.~.~A.}\ \bibnamefont
  {Jauregui}}, \bibinfo {author} {\bibfnamefont {M.}~\bibnamefont {Kayyalha}},
  \bibinfo {author} {\bibfnamefont {A.}~\bibnamefont {Kazakov}}, \bibinfo
  {author} {\bibfnamefont {I.}~\bibnamefont {Miotkowski}}, \bibinfo {author}
  {\bibfnamefont {L.~P.}\ \bibnamefont {Rokhinson}}, \ and\ \bibinfo {author}
  {\bibfnamefont {Y.~P.}\ \bibnamefont {Chen}},\ }\href@noop {} {\bibfield
  {journal} {\bibinfo  {journal} {App. Phys. Lett.}\ }\textbf {\bibinfo
  {volume} {112}},\ \bibinfo {pages} {093105} (\bibinfo {year}
  {2018})}\BibitemShut {NoStop}%
\bibitem [{\citenamefont {Fisher}\ and\ \citenamefont
  {Lee}(1981)}]{fisher1981relation}%
  \BibitemOpen
  \bibfield  {author} {\bibinfo {author} {\bibfnamefont {D.~S.}\ \bibnamefont
  {Fisher}}\ and\ \bibinfo {author} {\bibfnamefont {P.~A.}\ \bibnamefont
  {Lee}},\ }\href@noop {} {\bibfield  {journal} {\bibinfo  {journal} {Phys.
  Rev. B}\ }\textbf {\bibinfo {volume} {23}},\ \bibinfo {pages} {6851}
  (\bibinfo {year} {1981})}\BibitemShut {NoStop}%
\bibitem [{\citenamefont {Iida}\ \emph {et~al.}(1990)\citenamefont {Iida},
  \citenamefont {Weidenm{\"u}ller},\ and\ \citenamefont
  {Zuk}}]{iida1990statistical}%
  \BibitemOpen
  \bibfield  {author} {\bibinfo {author} {\bibfnamefont {S.}~\bibnamefont
  {Iida}}, \bibinfo {author} {\bibfnamefont {H.}~\bibnamefont
  {Weidenm{\"u}ller}}, \ and\ \bibinfo {author} {\bibfnamefont
  {J.}~\bibnamefont {Zuk}},\ }\href@noop {} {\bibfield  {journal} {\bibinfo
  {journal} {Ann. Phys.}\ }\textbf {\bibinfo {volume} {200}},\ \bibinfo {pages}
  {219} (\bibinfo {year} {1990})}\BibitemShut {NoStop}%
\bibitem [{\citenamefont {Haim}\ \emph {et~al.}(2015)\citenamefont {Haim},
  \citenamefont {Berg}, \citenamefont {von Oppen},\ and\ \citenamefont
  {Oreg}}]{haim2015signatures}%
  \BibitemOpen
  \bibfield  {author} {\bibinfo {author} {\bibfnamefont {A.}~\bibnamefont
  {Haim}}, \bibinfo {author} {\bibfnamefont {E.}~\bibnamefont {Berg}}, \bibinfo
  {author} {\bibfnamefont {F.}~\bibnamefont {von Oppen}}, \ and\ \bibinfo
  {author} {\bibfnamefont {Y.}~\bibnamefont {Oreg}},\ }\href@noop {} {\bibfield
   {journal} {\bibinfo  {journal} {Phys. Rev. Lett.}\ }\textbf {\bibinfo
  {volume} {114}},\ \bibinfo {pages} {166406} (\bibinfo {year}
  {2015})}\BibitemShut {NoStop}%
\bibitem [{\citenamefont {Ilan}\ \emph {et~al.}(2012)\citenamefont {Ilan},
  \citenamefont {Cayssol}, \citenamefont {Bardarson},\ and\ \citenamefont
  {Moore}}]{ilan2012nonequilibrium}%
  \BibitemOpen
  \bibfield  {author} {\bibinfo {author} {\bibfnamefont {R.}~\bibnamefont
  {Ilan}}, \bibinfo {author} {\bibfnamefont {J.}~\bibnamefont {Cayssol}},
  \bibinfo {author} {\bibfnamefont {J.~H.}\ \bibnamefont {Bardarson}}, \ and\
  \bibinfo {author} {\bibfnamefont {J.~E.}\ \bibnamefont {Moore}},\ }\href@noop
  {} {\bibfield  {journal} {\bibinfo  {journal} {Phys. Rev. Lett.}\ }\textbf
  {\bibinfo {volume} {109}},\ \bibinfo {pages} {216602} (\bibinfo {year}
  {2012})}\BibitemShut {NoStop}%
\bibitem [{\citenamefont {Zhang}\ \emph {et~al.}(2009)\citenamefont {Zhang},
  \citenamefont {Liu}, \citenamefont {Qi}, \citenamefont {Dai}, \citenamefont
  {Fang},\ and\ \citenamefont {Zhang}}]{zhang2009topological}%
  \BibitemOpen
  \bibfield  {author} {\bibinfo {author} {\bibfnamefont {H.}~\bibnamefont
  {Zhang}}, \bibinfo {author} {\bibfnamefont {C.-X.}\ \bibnamefont {Liu}},
  \bibinfo {author} {\bibfnamefont {X.-L.}\ \bibnamefont {Qi}}, \bibinfo
  {author} {\bibfnamefont {X.}~\bibnamefont {Dai}}, \bibinfo {author}
  {\bibfnamefont {Z.}~\bibnamefont {Fang}}, \ and\ \bibinfo {author}
  {\bibfnamefont {S.-C.}\ \bibnamefont {Zhang}},\ }\href@noop {} {\bibfield
  {journal} {\bibinfo  {journal} {Nat. Phys.}\ }\textbf {\bibinfo {volume}
  {5}},\ \bibinfo {pages} {438} (\bibinfo {year} {2009})}\BibitemShut {NoStop}%
\bibitem [{\citenamefont {Zhang}\ \emph {et~al.}(2012)\citenamefont {Zhang},
  \citenamefont {Kane},\ and\ \citenamefont {Mele}}]{PhysRevB.86.081303}%
  \BibitemOpen
  \bibfield  {author} {\bibinfo {author} {\bibfnamefont {F.}~\bibnamefont
  {Zhang}}, \bibinfo {author} {\bibfnamefont {C.~L.}\ \bibnamefont {Kane}}, \
  and\ \bibinfo {author} {\bibfnamefont {E.~J.}\ \bibnamefont {Mele}},\
  }\href@noop {} {\bibfield  {journal} {\bibinfo  {journal} {Phys. Rev. B}\
  }\textbf {\bibinfo {volume} {86}},\ \bibinfo {pages} {081303} (\bibinfo
  {year} {2012})}\BibitemShut {NoStop}%
\bibitem [{\citenamefont {Roy}\ \emph {et~al.}(2016)\citenamefont {Roy},
  \citenamefont {Roychowdhury},\ and\ \citenamefont {Das}}]{Roy_2016}%
  \BibitemOpen
  \bibfield  {author} {\bibinfo {author} {\bibfnamefont {S.}~\bibnamefont
  {Roy}}, \bibinfo {author} {\bibfnamefont {K.}~\bibnamefont {Roychowdhury}}, \
  and\ \bibinfo {author} {\bibfnamefont {S.}~\bibnamefont {Das}},\ }\href@noop
  {} {\bibfield  {journal} {\bibinfo  {journal} {New Journal of Physics}\
  }\textbf {\bibinfo {volume} {18}},\ \bibinfo {pages} {073038} (\bibinfo
  {year} {2016})}\BibitemShut {NoStop}%
\end{thebibliography}%

\end{document}